\documentclass{aastex}
\usepackage{emulateapj5}

\begin{document}


\title{Spectral Resolution-linked Bias in Transit Spectroscopy of Extrasolar Planets}



\author{Drake Deming\altaffilmark{1,2} and Kyle Sheppard\altaffilmark{1}}

\altaffiltext{1}{Department of Astronomy, University of Maryland at College Park, 
College Park, MD 20742, USA}
\altaffiltext{2}{NASA Astrobiology Institute's Virtual Planetary Laboratory}


\begin{abstract}
We re-visit the principles of transmission spectroscopy for transiting
extrasolar planets, focusing on the overlap between the planetary
spectrum and the illuminating stellar spectrum. Virtually all current
models of exoplanetary transmission spectra utilize an approximation
that is inaccurate when the spectrum of the illuminating star has a
complex line structure, such as molecular bands in M-dwarf spectra.
In those cases, it is desirable to model the observations using a
coupled stellar-planetary radiative transfer model calculated at high
spectral resolving power, followed by convolution to the observed
resolution.  Not consistently accounting for overlap of stellar
M-dwarf and planetary lines at high spectral resolution can bias the
modeled amplitude of the exoplanetary transmission spectrum, producing
modeled absorption that is too strong.  We illustrate this bias using
the exoplanet TRAPPIST-1b, as observed using HST/WFC3.  The bias in
this case is about 250 parts-per-million, 12\% of the modeled
transit absorption.  Transit spectroscopy using JWST will have access
to longer wavelengths where the water bands are intrinsically
stronger, and the observed signal-to-noise ratios will be higher than
currently possible.  We therefore expect that this resolution-linked
bias will be especially important for future JWST observations of
TESS-discovered super-Earths and mini-Neptunes transiting M-dwarfs.

\end{abstract}




\section{Introduction}  \label{sec:intro}

Transit spectroscopy is a productive method to derive the atmospheric
properties of transiting extrasolar planets.  The first successful
transit spectroscopy exploited strong atomic lines to probe the
exoplanetary atmosphere (e.g.,\citealp{charb02, redfield}). In the
past few years, the WFC3 instrument on the Hubble Space Telescope has
been used to measure water vapor absorption at 1.4\,$\mu$m in giant
planets \citep{deming13, kreidberg1, fraine14, nikolov15, evans16,
  sing16}, and provide stringent upper limits for smaller planets
\citep{kreidberg2, knutson}.

A future goal of transit spectroscopy is to measure molecular
absorption in the atmosphere of a habitable super-Earth transiting a
nearby M-dwarf star, e.g. using JWST \citep{deming09, cowan}.  The
TESS mission \citep{ricker} will discover the best systems for this
quest, but already the ground-based surveys are finding excellent
candidates such as GJ1132b \citep{berta}, TRAPPIST-1 \citep{gillon,
  dewit, barstow, gillon17}, and even a (non-transiting) planet orbiting Proxima
Centauri \citep{proxima}. Considering the increasing attention paid to
planets that orbit M-dwarf stars, and the imminent plans for JWST
observations \citep{stevenson, greene}, factors that affect transit
spectroscopy of such planets are of urgent interest.

In this {\it Letter} we describe an important effect that is relevant
to transit spectoscopy of planets transiting M-dwarf stars. In brief,
the line absorption spectrum of the host star can bias the inferred
transit spectrum of the planet. The bias is not in the observations
per se, it occurs when modeling of the transit spectrum utilizes
insufficient spectral resolution.  We call this effect
resolution-linked bias (hereafter, RLB).  Sec.~2 describes the source
of RLB, and Sec.~3 evaluates its magnitude for the recent case of the
TRAPPIST-1b exoplanet.  Sec.~4 concludes with some remarks on the
future importance of the RLB effect.

\section{Source of Resolution-linked Bias}

We begin here by reminding the reader of the spectral resolving power
needed to resolve intrinsic line widths in stellar and planetary
spectra (Sec.~2.1).  We then explain the source of RLB using a simple
numerical example (Sec.~2.2) and physical arguments, and then with a
general mathematical formalism (Sec.~2.3).

\subsection{Resolving Stellar and Exoplanetary Spectra}

Two principal mechanisms broaden lines in stellar and planetary
spectra: Doppler broadening and pressure broadening. The physical
principles of spectral line broadening are already well understood
\citep{rybicki}, but we apply these principles to the TRAPPIST-1
system in order to provide a context for our discussion of RLB.

The Doppler width of a spectral line is determined by the
line-of-sight thermal motion of the absorbing species, being
proportional to the square root of temperature ($\Delta\nu = \nu_{0}/c
\sqrt(2kT/m)$), where $k$ is Boltzmann's constant and $m$ is the mass
of the absorbing atom or molecule).  Table~1 summarizes the Doppler
and pressure-broadened widths for 1.4\,$\mu$m water vapor absorption in the continuum and
line-forming regions of the stellar and planetary atmospheres in the
TRAPPIST-1b system. We adopted pressures and temperatures for these
regions based on the Teff/$\log g$/[M/H] = 2500/5.5/0.0 Phoenix stellar model
\citep{husser}, and an isothermal atmosphere for the planet at the
estimated equilibrium temperature of 400K \citep{gillon17}.  The exact
pressures and temperatures are not critical to the main point of
Table~1: nearly all stellar and exoplanetary observed spectra fail to
resolve the widths of individual spectral lines.  The requisite
spectral resolving powers ($\nu/\Delta\nu$) range from over $10^5$ for
the continuum-forming regions of the stellar atmosphere, to over
$10^7$ for low pressure regions of the exoplanetary atmosphere.

No current or planned space-borne spectrometer can attain the spectral
resolving powers listed in Table~1.  Grism spectroscopy with HST/WFC3
at 1.4\,$\mu$m achieves a resolving power of about 130 (and usually
less because of binning), which is several orders of magnitude less
than required to resolve individual spectral lines of water vapor.
The situation is similar for JWST, where near- and mid-IR resolving
powers will range from 700 (NIRISS) to 2700 (NIRSpec).  We do not make
this comparison as a criticism of these instruments: low to moderate
spectral resolving power is necessary in order to attain adequate
signal-to-noise for many measurements.  Rather, we emphasize that the
observed spectra do not come even close to resolving the intrinsic
line widths.

Figure~1 shows an example of an M-dwarf spectrum, from the Phoenix
model \citep{husser} appropriate for the host star in TRAPPIST-1.  The
spectrum at a resolving power ($\lambda/\delta\lambda$) of $5.0
  \times 10^5$ is a forest of line-resolved structure (in gray on top panel).
We convolved that spectrum with a Gaussian instrument profile, to a
spectral resolving power of 1000 (blue spectrum in the top panel).  At
that resolving power, water absorption remains prominent in the
spectrum, but the depths of the bands appear modest.  However, at full
resolution the individual absorption lines due to water vapor are very
strong: many have core intensities below 0.1 of the local continuum
level.  Note also that these `lines' are not individual quantum
transitions.  Between 7000 and 9000 $cm^{-1}$ there are about 50
million discrete transitions in the line list from
\citet{barber}. What appear to be single lines on Figure~1 are
actually composed of (in most cases) hundreds of distinct quantum
transitions.  However, the wavelength scale of structure in the
spectrum is still dictated by the broadening mechanisms discussed
above.

In addition to line broadening, relative wavelength shifts of lines in
the star versus a transiting planet are relevant to RLB.  Many
transiting planets have circular orbits, providing only minimal
Doppler shifts between the stellar and planetary lines during transit
(e.g., from stellar rotation, and the projection of the planet's
  orbital velocity onto the stellar disk).  Line overlap due to small
relative Doppler shifts tend to maximize RLB (Sec.~3). TRAPPIST-1 is a
rapid rotator (P=1.4d, \citealp{gillon}), with an equatorial rotation
speed of 4.2 km/sec.  The planet's orbital velocity is 80 km/sec,
  and the component projected onto the stellar disk during transit is
  $\leq 4$ km/sec, and that tends to reduce the relative radial velocity
  difference, assuming that the planet orbits in the same direction as
  stellar rotation. A significant RLB effect results from the
velocity difference, as we show in Sec.~3.

Stellar granular convection is another source of Doppler shifts. IR
lines in M-dwarf spectra are not well studied in terms of convective
Doppler shifts, but hydrodynamic simulations indicate granular
convective velocities of $\sim 240$\,meters/sec \citep{ludwig}, too
small to be significant in this context.

Winds in hot exoplanet atmospheres can produce Doppler shifts of
planetary lines by several km/sec \citep{kempton}, but a velocity of
that magnitude will still leave substantial overlap between the
stellar and planetary lines.

The mechanisms discussed above do not shift exoplanetary spectral
lines enough to prevent substantial overlap with stellar lines, and
the overlap is the source of RLB.

\subsection{A Simple Numerical Example}

We begin our explanation of RLB with a simple numerical example.
Consider wavelengths $\lambda_1$ and $\lambda_2$, and out-of-transit
stellar fluxes $F_1$ and $F_2$.  The fine-scale spectral structure
seen in Figure~1 can lead to a significant difference between $F_1$
and $F_2$, so (for example) we might have $F_2 = 0.1F_1$ if
$\lambda_2$ is centered on a water vapor line in both the planet and
star - which is possible because of the small relative Doppler shift
as noted above.  We let the area (i.e., solid angle) of the planet (in
units of the stellar disk area) at wavelength $\lambda_1$ be $A$, and
let the absorbing atmospheric annulus have area $\delta$, only at
wavelength $\lambda_2$.

Now consider the depth of the transit, by comparing stellar fluxes
in transit and out of transit.  At wavelength $\lambda_2$ where the
exoplanetary atmosphere is strongly absorbing, the in transit flux is
$0.1F_1(1-A-\delta)$, so the depth of the transit being (out-in)/out, is:

\begin{equation}
(0.1F_1-0.1F_1(1-A-\delta))/0.1F_1 = A+\delta
\end{equation}
and similarly at $\lambda_1$, the depth of the transit is $A$.
As long as $\lambda_1$ and $\lambda_2$ are spectrally resolved, there
is no RLB effect.  However, as we noted in Sec.~2.1, fine scale
structure in M-dwarf spectra is often unresolved and observations
can only sense the total flux in a resolution element.  So the
observed transit depth becomes:

\begin{equation}
(1.1F_1-(F_1(1-A)+0.1F_1(1-A-\delta)))/1.1F_1 
= A+0.091\delta
\end{equation}

The transit light curve at $\lambda_2$ is diluted by the greater
stellar flux at $\lambda_1$, where neither the planet nor star has
water vapor absorption. Therefore the transit at the observed spectral
resolving power can in principle exhibit a much lower apparent absorption than
actually occurs in the monochromatic spectrum.

In this example, the RLB effect would be absent if the two wavelengths
were spectrally resolved, because the ratio of in-transit to
out-of-transit flux is accurate at each wavelength.  However, when the
wavelengths are averaged by low spectral resolving power, then the ratio
of the average fluxes does not preserve the spectral transmission
information accurately. Fundamentally that derives from the
nonlinearity of ratios: a ratio of averages is not equal to an average
of ratios.

Physically, we can envision a case where water absorption in the
exoplanetary atmosphere occurs in a series of discrete lines that are
centered on stellar water absorption lines.  At the water wavelengths
the star is much fainter than in the continuum, so the water
absorption during transit is readily overwhelmed by flux at
wavelengths where no water absorption occurs.  In a real situation,
important additional factors include the relative strengths of water
lines in the planet and star due to their different excitation
temperatures, different line broadening, and the stellar rotational
velocity.  For example, if the planetary lines were much broader than
the stellar lines, or the planet's orbital velocity was large and
retrograde, then the RLB effect will be correspondingly reduced by the
mismatch in line profiles or wavelengths.  We include all relevant
effects in our calculations of Sec.~3.

\subsection{Mathematical Formalism}

We here give a more formal mathematical description of RLB.  Let the
monochromatic flux of the star at wavelength $\lambda$ be
$F_{\lambda}$, and let the absorption of the exoplanetary atmosphere
at $\lambda$ be $A_{\lambda}$ ($1-A_{\lambda}$ is transmitted).  We
observe the spectrum with less than line-resolved resolution, and we
denote the spectral instrumental profile as $I_{\lambda}$.  The
measured spectrum during a transit is of the form $(out-in)/out$ where
$in$ and $out$ refer to the measured flux in and out of transit.  In
that form, the measured signal $S_{\lambda}$ is:

\begin{equation}
S_{\lambda} = \frac{I_{\lambda} \otimes F_{\lambda} - I_{\lambda} \otimes (F_{\lambda} (1-A_{\lambda}))}
{I_{\lambda} \otimes F_{\lambda}},
\end{equation}

\begin{equation}
= 1-\frac{I_{\lambda} \otimes (F_{\lambda} (1-A_{\lambda}))}
{I_{\lambda} \otimes F_{\lambda}},
\end{equation}

where $\otimes$ denotes convolution.  Instead of Eq.(4), a more usual
practice is to represent the star out of transit using a low
resolution spectrum $F^{'}_{\lambda}$, and to write the in transit
flux as $F^{'}_{\lambda} (1-A_{\lambda})$.  In that case, the stellar
flux cancels in the $(out-in)/out$ expression, and $S_{\lambda} =
A_{\lambda}$.  But since $A_{\lambda}$ is often modeled at a spectral
resolution higher than the observations, the result is convolved with
$I_{\lambda}$ in an {\it ad hoc} fashion, i.e.;

\begin{equation}
S_{\lambda} = I_{\lambda} \otimes A_{\lambda}
\end{equation}.

The necessity for doing a convolution to the observed resolution in an
{\it ad hoc} fashion is a clue that Eq.(5) is not strictly correct.
In Sec.~3, we illustrate the difference between the approximation
represented by Eq.(5) and the strictly correct Eq.(4); the magnitude
of that difference is the RLB effect.

\section{Magnitude of Resolution-linked Bias for a Representative Case}

We here calculate the RLB effect for the exoplanet TRAPPIST-1b
\citep{dewit}.  We choose this system as an example because the host
star is a late M-dwarf, and its spectrum is rich in molecular
absorption lines, potentially producing a significant RLB
effect. Sec.~3.1 describes the radiative transfer aspect of our
calculations, Sec.~3.2 explains our treatment of line opacity, and
Sec.~3.3 describes how we tested our codes.  Sec.~3.4 shows the
magnitude of the RLB effect for TRAPPIST-1b.

\subsection{Exoplanetary and Stellar Model Spectra}

We model the star using a  Teff/$\log\,g$/[M/H] = 2500/5.5/0.0 Phoenix stellar model
\citep{husser}, and we use an isothermal atmosphere at 400K for the
planet (as in Sec.~2.1). We derive our water absorption line opacities
for both the planet and star using the line list from
\citet{barber}. Our transit spectrum of the TRAPPIST-1b planet is
calculated by the code described by \citet{deming13}; our stellar
spectral code is new to this paper.  We produce both the flux and
intensity spectrum of the TRAPPIST-1 star, based on the
pressure-temperature relation of Phoenix model cited above.  Because
the planet is smaller than the stellar disk, disk-resolved intensity
(not flux) calculations are appropriate for modeling the RLB effect.
However, the Phoenix models tabulate emergent flux, so it is
convenient to check our code versus the Phoenix flux spectrum.  We
adopt the pressure-temperature relation, and the solar composition and
$log(g)=5.5$, of the Phoenix model.  Our continuous opacity is due
to H$_2$ Rayleigh scattering \citep{dalgarno}, as well as H$_2$
collision-induced opacity \citep{zheng, borysow}. For both the planet
and star, our line opacity uses only water vapor. We calculate the
emergent intensity using a high-resolution wavelength grid ($0.4$ km
sec$^{-1}$ spacing). At each wavelength, we interpolate the optical
depths in the model to a uniform grid in $log(\tau)$ to facilitate
accurate integration.  Adopting an LTE source function, we calculate
emergent intensity from the formal solution of the transfer equation
\citep{rybicki}, integrating on the uniform grid using Laguerre
polynomial weights.  We derive the flux spectrum using an 8-point
Gaussian quadrature integration over intensity as a function of
emergent angle.

\subsection{Line Opacity Calculations}

The water vapor line list from \citet{barber} is too extensive to
treat each line with an individual Voigt profile.  As advocated by
\citet{grimm}, we here describe exactly how we calculate line
opacity. Our method is based on the fact that many line centers
overlap within each Doppler width, so we average those lines to form a
single equivalent 'pseudo-line'.  We adopt a grid of equally spaced
frequencies (this is different than the grid used to calculate the
spectra).  Within each grid interval we add the strengths of all lines
whose central rest frequency falls in the interval to form a
  single pseudo-line per interval.  Because the lines have
different lower state excitation energies, the sum of their strengths
is temperature-dependent.  We tabulate the summed strength at 10
temperatures spanning the range in the models, and we fit the results
to within 2\% in line strength using a 4th order polynomial as a
  function of temperature.  This pre-processing produces a list of
pseudo-lines spaced at 0.005 cm$^{-1}$ intervals, and reduces the
number of lines by a factor of $\sim 100$ from the original
\citet{barber} list. We use a pressure broadening coefficient of 0.06
cm$^{-1}$ per atmosphere for all lines. \citet{husser} infer
  negligibly small microturbulence for TRAPPIST-1, so we ignore
  microturbulence except for one exploratory calculation mentioned in
  Sec.~3.4. 

\subsection{Tests of the Calculations}

To ensure accurate evaluation of the RLB effect, we tested the
precision and accuracy of our calculations. Our planetary
transmittance code was previously checked against independent
calculations from \citet{fortney10} (see Figure~12 of the
\citealp{deming13} paper).  The same code was also compared to a
second independent calculation by \citet{line}.  In our stellar
calculations, we verified adequate precision of the integrations by
setting the atmospheric temperature to be isothermal, and comparing
the calculated intensity and flux to the analytic Planck function at
that temperature (an LTE isothermal atmosphere emits as a blackbody).
We found that the emergent intensities were accurate to $0.1\%$, and
the flux was accurate to $1.1\%$.  (Note that we do not require
accuracy of the intensities and fluxes in an absolute sense, we only
require that the relative shape of emergent spectral lines be
correct.) Finally, we tested our flux calculation by
comparison to the high resolution Phoenix flux from \citet{husser}.
That comparison is shown for an expanded region near the 1.4\,$\mu$m
water band in Figure~2.  Overall, we find excellent agreement, with
our spectrum showing more absorption in some regions.  Those
differences are as expected, because the Phoenix line opacity uses a
line strength cutoff (P. Hauschildt, private communication), whereas
we include all of the lines from \citet{barber}.  We conclude that our
calculations are accurate, and - most important - they are
self-consistent from planet to star.

We tested our line averaging procedure by varying the size of the grid
interval used for the averaging. In the limit where the interval
approaches zero, our averaging reduces to a true line-by-line
calculation.  We verified that the spectrum approaches that limit as
we reduced the interval.  We found that interval widths less than
0.01 cm$^{-1}$ produced negligible differences from a line-by line
calculation, but to be conservative we used a 0.005 cm$^{-1}$ grid for
our models. 

\subsection{Magnitude of the Effect for TRAPPIST-1b}

We calculated a transmission spectrum for TRAPPIST-1b, adopting an
isothermal solar composition atmosphere at T=$400$K.  In order
to explore the potential temperature dependence of RLB, we also
calculated for an exoplanetary temperature of $800$K (although the
planet is not sufficiently irradiated to be that hot).  We did an
exact calculation using Eq. (4), and also the conventional (i.e.,
approximate) calculation using Eq. (5).  In all cases we used the
pressure-temperature relation of the Phoenix model to calculate
emergent intensity spectra, as described in Sec.~3. We multiplied each
intensity spectrum times the planetary transmittance, and averaged
over 11 points equally spaced across the stellar disk during transit.
In the exact calculation, we accounted for the stellar rotation, the
projection of the planet's orbital velocity, and the gravitational
redshift when multiplying the stellar spectra times the transmittance
of the exoplanetary atmosphere.

Figure 3 shows the results of our calculations.  The transit spectrum
of a solar composition atmosphere for TRAPPIST-1b in this HST/WFC3
bandpass has an amplitude of about 2000 ppm (as already shown by
\citealp{dewit}). The difference between the exact and approximate
calculation represents the RLB effect. RLB is significant in the
strongest portion of the band between 1.35 and 1.4\,$\mu$m.  The
difference (plotted on the lower portion of Figure~3) is about
  250 ppm, 12\% of the peak transmittance. For the 800K
atmosphere, the RLB effect is about 900 ppm, almost four times as
  large as for the 400K atmosphere.  A factor of two is due to
  the greater scale height of the 800K atmosphere, but a comparable factor
  comes from the excitation structure of the water band, wherein the
  higher planetary temperature increases the match between the stellar
  and exoplanetary spectrum. We also tested the effect of a possible 2
  km/sec stellar microturbulence, and we find that it decreases the
  RLB effect only slightly (35 ppm).

\section{Concluding Remarks}

The RLB effect for TRAPPIST-1b is not sufficient to alter the
conclusions of the study by \citet{dewit}.  Interestingly, the effect
is about the same amplitude as the {\it total} absorption due to water
vapor in a typical hot Jupiter (e.g., \citealp{deming13}). As for hot
Jupiters, most of them transit FGK stars, where the RLB effect due to
water vapor will be negligible.

Although RLB is not a major factor affecting current observations, it
will be important to include it when modeling future JWST
observations, especially of super-Earths and mini-Neptunes transiting
M-dwarfs.  TESS will discover many such worlds that are favorable for
observations using JWST.  Based on our exploratory calculations to
date, we expect that RLB will be strongest for hot planets in prograde
orbits around M-dwarf host stars, although it will be reduced for
cases where the planet's orbital velocity is very high.  The molecular
bands in the JWST spectral range tend to be intrinsically stronger
than in the HST/WFC3 band, and that should also lead to a larger RLB
effect.  Given anticipated high signal-to-noise from JWST transit
spectroscopy, it is likely that RLB will be well above the
observational precision in many cases. Because RLB affects the
strongest portions of molecular bands, ignoring it would cause
systematic errors in atmospheric retrievals at the highest
altitudes. Although it is primarily of concern for M-dwarf host stars,
stars as hot as the Sun can have significant absorption in the
fundamental carbon monoxide band near 4.6\,$\mu$m \citep{uitenbroek}.
Hence JWST spectroscopy of planets transiting GKM stars should be
evaluated for significant RLB effect on carbon monoxide retrievals.

\acknowledgements We thank an anonymous referee for insghtful and
  constructive comments that improved this paper, and we acknowledge
  informative correspondence with the Phoenix group, specifically
  Tim-Oliver Husser, Ansgar Reiners, and Peter Hauschildt.  We also
  gratefully acknowledge NASA grant NNX15AE19G, and support from
  NASA's Virtual Planetary Laboratory.







\clearpage

\begin{table}
\begin{center}
\caption{Doppler width, and pressure-broadened widths (in cm$^{-1}$)
  for water vapor lines near 1.4\,$\mu$m in the atmospheres of
  the star and planet 'b' in the TRAPPIST-1 system. \label{tbl-1}}
\begin{tabular}{llll}
\tableline\tableline
Atmospheric level & Temperature or Pressure & $\Delta\nu$ & $\nu/\Delta\nu$  \\
\tableline
 Star, continuum &  2760\,K    &  $0.038$   &  $1.89 \times 10^5$ \\
 Star, continuum &  1.75\,bar  &  $0.105$   &  $6.80 \times 10^4$ \\
 Star, line region  & 1700\,K  & $0.030$    &  $2.40 \times 10^5$ \\
 Star, line region  & 80\,mbar & $0.0048$    &  $1.50 \times 10^6$ \\
 Planet, troposphere & 400\,K  & $0.014$  & $ 4.95 \times 10^5$   \\
 Planet, troposphere & 1\,bar  & $0.060$ & $1.19 \times 10^5$   \\
 Planet, upper atmosphere & 10\,mbar  & $0.0006$ & $1.19 \times 10^7$   \\
 \tableline
\end{tabular}
\end{center}
\end{table}

\clearpage
\begin{figure}
  \epsscale{0.7} \plotone{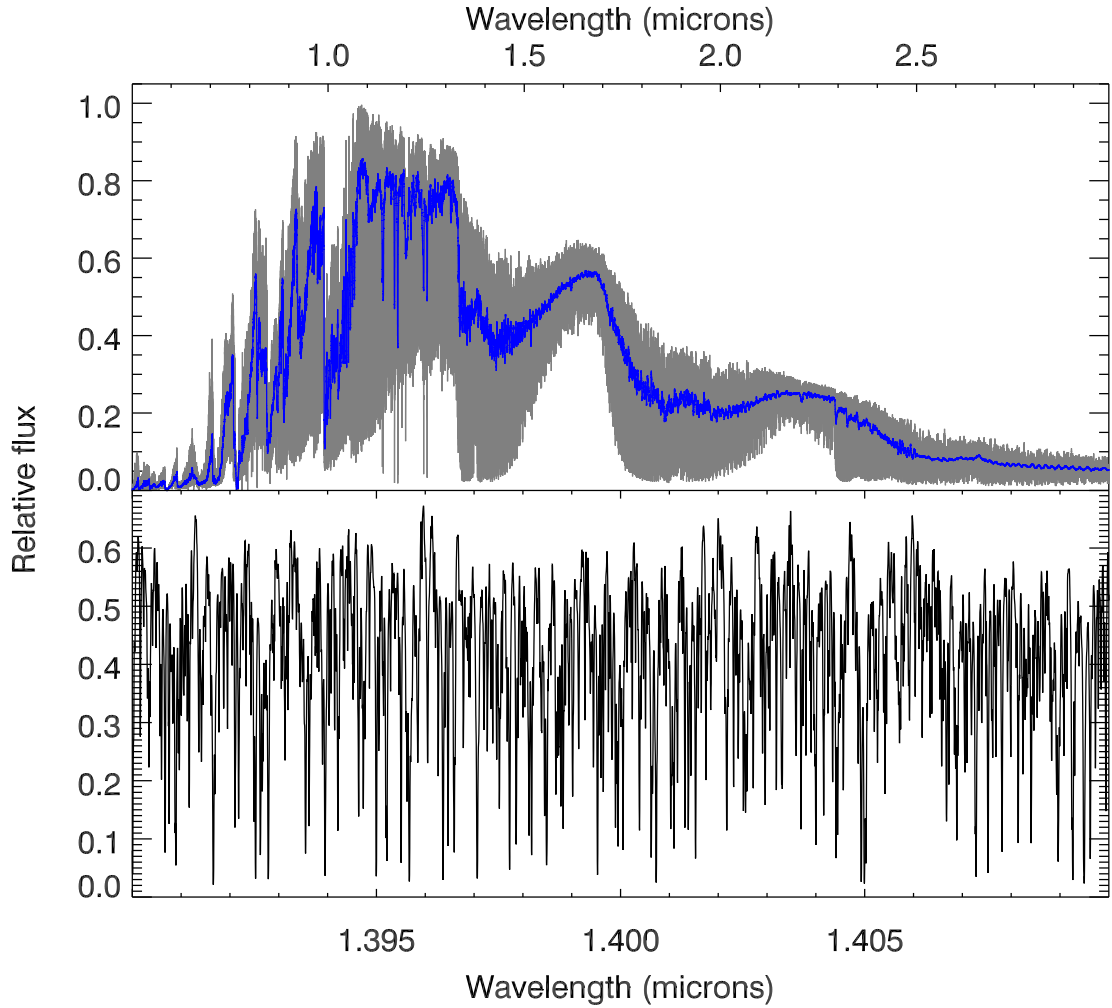}
\vspace{0.8in} 
\caption{Near-infrared flux from a Phoenix model atmosphere with
  Teff/log\,$g$/[Fe/H] of 2500/5.5/0.0 from \citet{husser}.  The top
  panel plots the spectrum at full resolution in gray, and convolved
  to a spectral resolving power of 1000 (blue).  The lower panel
  expands a small section of the spectrum near 1.4\,$\mu$m.  Note the
  deep and abundant absorption lines.}
\label{Fig1}
\end{figure}

\clearpage
\begin{figure}
  \epsscale{0.8} \plotone{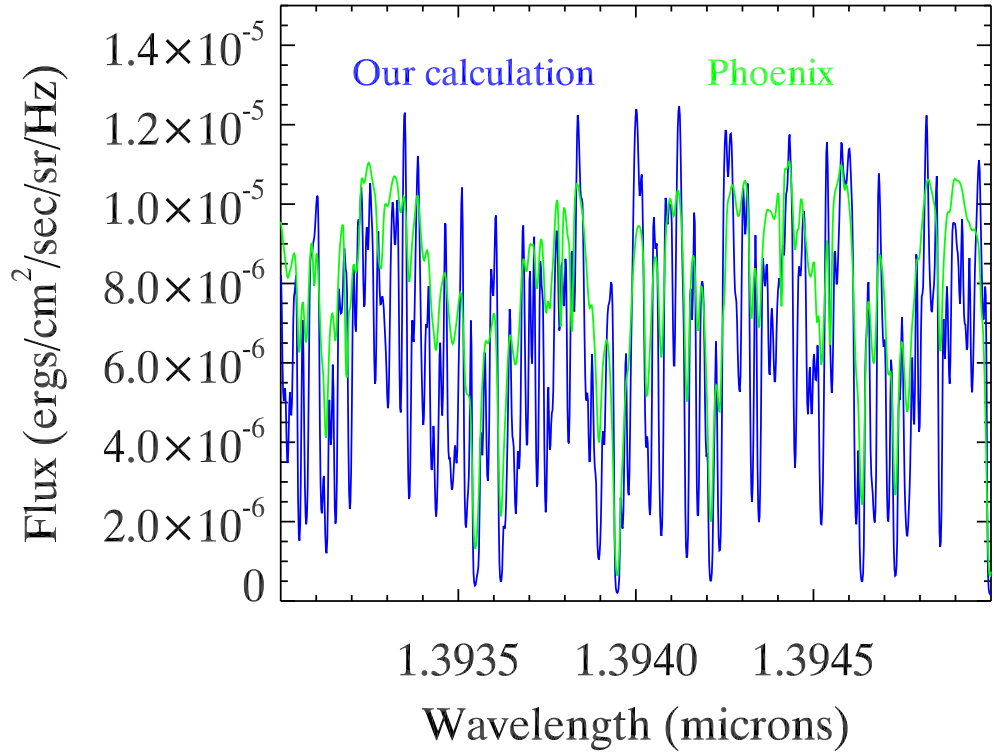}
\vspace{0.8in} 
\caption{For an expanded wavelength region near the 1.4\,$\mu$m water vapor band
  head, we here test our calculated stellar flux (blue line) by comparing to the
  flux from \citealp{husser} (green line). We use the temperature/pressure relation of
  the Phoenix model atmosphere with Teff/log\,$g$/[Fe/H] of
  2500/5.5/0.0.}
\label{Fig2}
\end{figure}

\clearpage
\begin{figure}
  \epsscale{0.6} \plotone{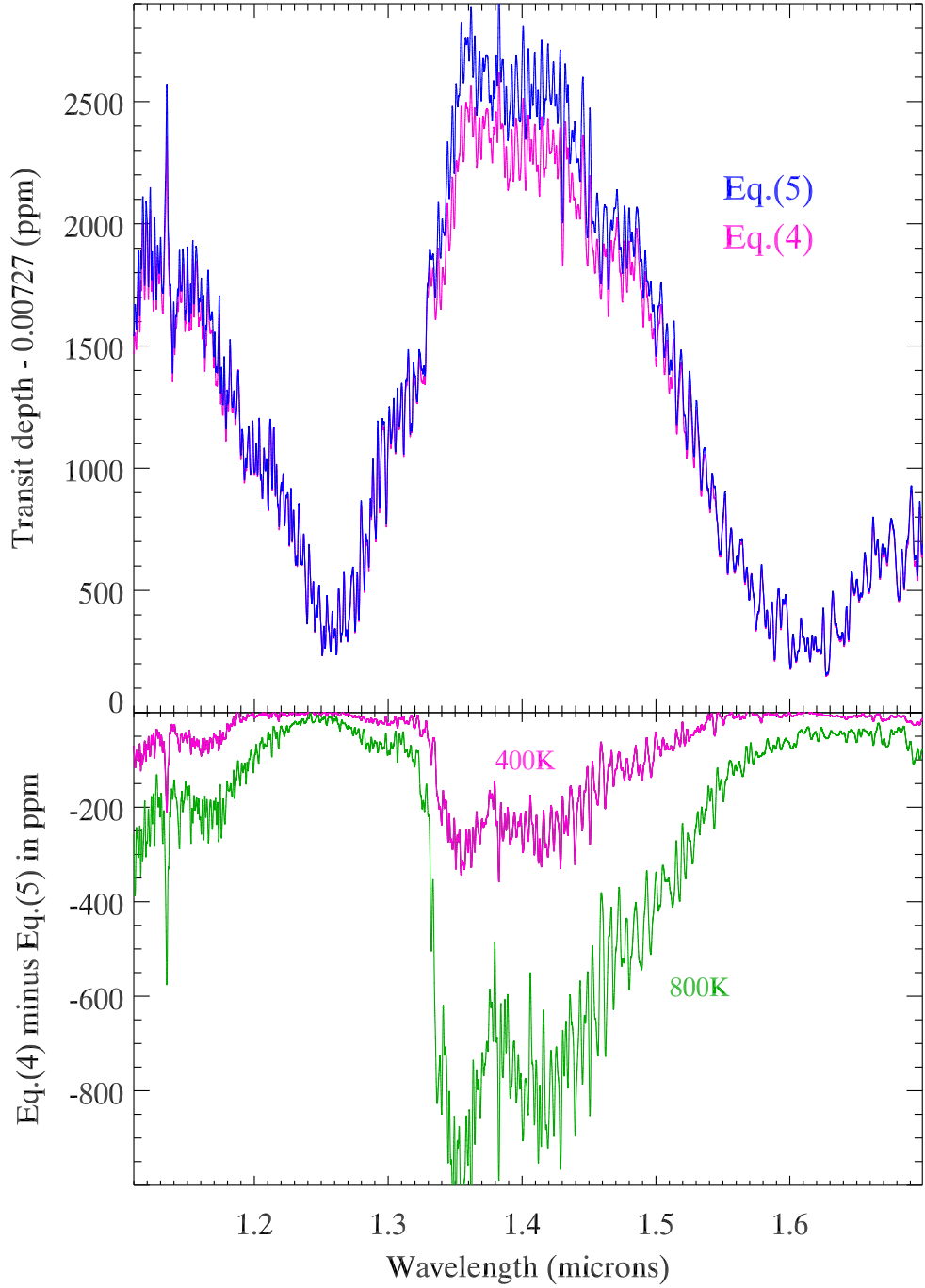}
\vspace{0.8in} 
\caption{ Top panel: transmission spectrum of TRAPPIST-1b modeled with
  the exact relation (Eq. 4) versus the approximate method that is
  commonly used (Eq. 5). Bottom panel: the difference between Eq.~4
  and Eq.~5 for our nominal model with an exoplanetary atmospheric
  temperature of $400$K. Also plotted is the difference if
  the exoplanetary temperature were 800K.}
\label{Fig3}
\end{figure}

\end{document}